\documentclass[preprint,preprintnumbers,suprescriptaddress,amsmath,amssymb,tightenlines]{revtex4}

\pdfoutput=1
\usepackage{verbatim,graphics,graphicx,color,slashed}
\usepackage{color}
\usepackage{gensymb}
\usepackage[normalem]{ulem}
\begin{document}


\title{Implication of the ALEPH 30 GeV dimuon resonance at the LHC}


\author{P. Ko}
\email[]{pko@kias.re.kr}
\affiliation{School of Physics, KIAS, Seoul 02455, Korea}
\affiliation{Quantum Universe Center, KIAS, Seoul 02455, Korea}

\author{Jinmian Li}
\email[]{jmli@kias.re.kr}
\affiliation{School of Physics, KIAS, Seoul 02455, Korea}

\author{Chaehyun Yu}
\email[]{chyu@korea.ac.kr}
\affiliation{Department of Physics, Korea University, Seoul 02841, Korea}


\date{\today}

\begin{abstract}
Recent reanalysis of ALEPH data on $Z\rightarrow b\bar{b} + X$  seems to indicate an existence 
of the dimuon excess around 30 GeV  with a branching fraction for 
$Z\rightarrow b\bar{b} \mu^+ \mu^-$ around $1.1 \times10^{-5}$.
In this letter, we discuss three different types of simplified models for this possible excess. 
In the first class of models,  we assume a new resonance couples to both $b\bar{b}$ and 
$\mu^+ \mu^-$. Within the allowed parameter space for the ALEPH data,  this type of models is excluded 
because of too large Drell-Yan production of dimuon from the $b\bar{b}$ collision 
at the LHC.
In the second model, we assume that the 30 GeV excess is a new gauge boson $Z'$ that couples 
to the SM $b$ and a new vectorlike singlet $B$ quark heavier than $Z$ and not to $b\bar{b}$.
Then one can account for the ALEPH data without conflict with the DY constraint. 
The new vectorlike quark $B$  can be pair produced at the LHC 8/13 TeV  by QCD with 
$\sigma (BB)\sim O(100-1000)$ pb, 
and $Bq$ production rate is $\sigma (Bq) \sim$ a few pb which is larger than $\sigma(Bb)$ 
roughly by an order of magnitude. 
Their signatures at the LHC would be $2b + 4\mu$, $b j + 2 \mu$ and $2b+2\mu$, respectively, 
which however might have been excluded already by  LHC run I and II data since the multi-muon events 
have low SM background and are rare at the LHC.
In the third model, we consider $Z\rightarrow Z' \phi$ followed by $Z' \rightarrow \mu^+ \mu^-$ and 
$\phi \rightarrow b\bar{b}$ assuming that the Higgs field for $Z'$ mass is also charged under the SM 
$U(1)_Y$ gauge symmetry.   In this class of model, we could accommodate the 
$\textrm{Br}(Z\rightarrow b\bar{b} \mu^+ \mu^-) \sim 1.1 \times 10^{-5}$  
if we assume very large $U(1)'$ charge for the $U(1)'$ breaking Higgs field.  
Finally, we study various kinematic distributions of muons and $b$ jets in all the three models, 
and  find that none of the models we consider in this paper are not compatible with the  kinematic 
distributions extracted from  the ALEPH data.
\end{abstract}


\maketitle

\section{Introduction}

A recent reanalysis of the archived  ALEPH data on $Z \rightarrow b \bar{b} +X$ might suggest   
an interesting possibility of a new resonance $X$ with mass 30 GeV  decaying into 
$\mu^+ \mu^-$~\cite{aleph30}:
\begin{eqnarray}
\textrm{Br}( Z \rightarrow b \bar{b} \mu^+ \mu^- )  & \sim & 1.1 \times 10^{-5},
\label{brbbmm}%
\\
\Gamma_{\rm tot} (X) & = & (1.78 \pm 1.14) ~{\rm GeV} .
\end{eqnarray}
Dielectron channel also shows some excess, which however is less prominent than the dimuon 
channel.  
The $\cos\theta_\mu^*$ distribution of a muon in the rest frame of dimuon system 
with respect to the direction of the dimuon system in the rest frame of $Z$ 
shows peaks around 
$\cos\theta_\mu^* \approx \pm 1$, which would prefers $X$ being a spin-1 particle.

In addition, a few interesting kinematic distributions are presented in Ref.~\cite{aleph30}.
In the signal region, the minimum angle between a muon and 
the leading $b$ jet is within $15\degree$
and the angle of the other muon-jet combination is in the range of
$5\degree$ to $20\degree$. Also the relative transverse momentum distribution
of the closest muon-jet pair is smaller than $4$ GeV~\cite{aleph30}.   These distributions have to be 
reproduced by any working models for the 30 GeV dimuon excess. 

\begin{figure}[thb] 
\includegraphics[width=0.3\textwidth]{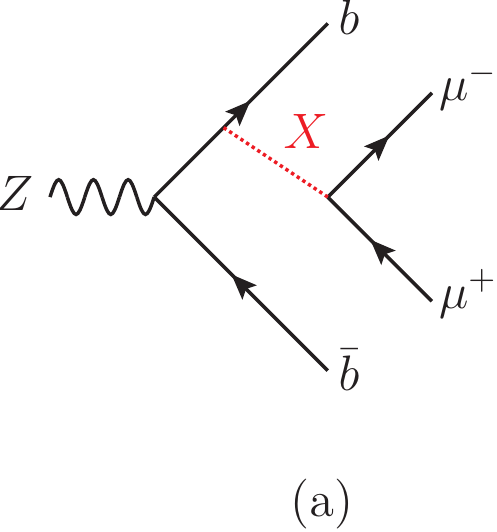}
\includegraphics[width=0.3\textwidth]{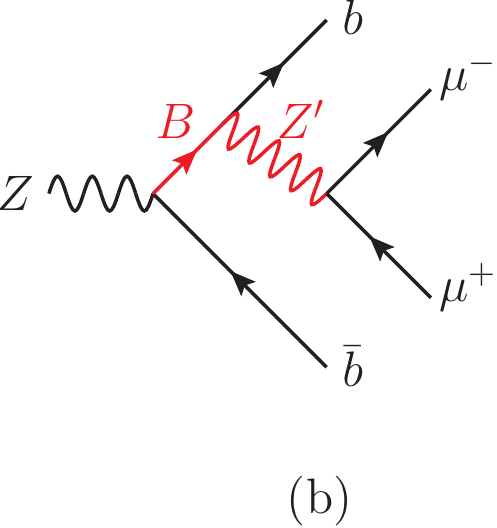}
\includegraphics[width=0.3\textwidth]{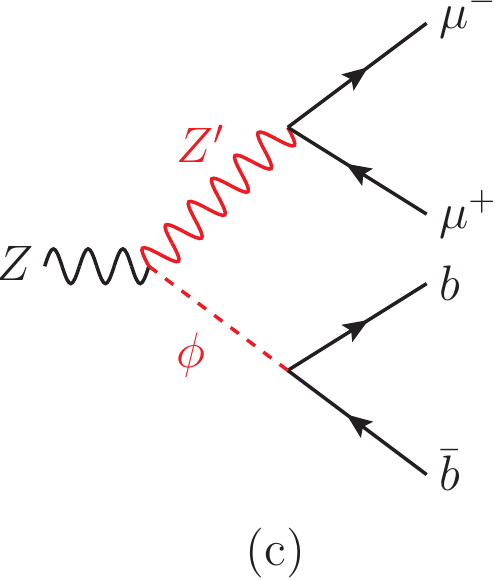}
\caption{\label{fig:diag} 
Feynman diagrams for the $Z\to b\bar{b}\mu^+\mu^-$ decay.
}
\end{figure}

In this letter, we consider three types of simplified models for this 
30 GeV dimuon excess, for which the relevant Feynman diagrams are shown in 
Fig.~\ref{fig:diag} (a)--(c).
Note that these three types of Feynman diagrams exhaust all possible  tree-level mechanisms 
for this dimuon  excess.  

In Sec.~\ref{sec:model1}, we assume a new resonance $X$ couples to both $b\bar{b}$ 
and $\mu^+\mu^-$ with $X$ being  (pseudo)scalar or (axial) vector boson, as shown in 
Fig.~\ref{fig:diag}(a).  Then we perform comprehensive phenomenological 
study on $X \rightarrow b \bar{b} \mu^+ \mu^-$ and related  processes such as 
$X \rightarrow 4 \mu, 4 b$, finding out the parameter space that can 
account for the ALEPH data.  Then within this parameter space, we study the predictions 
involving $X$  at the LHC:  Drell-Yan (DY) process from 
$b\bar{b} \rightarrow X \rightarrow \mu^+ \mu^-$, and $X$ productions in 
$ t \rightarrow b W X$,  $b\bar{b} X$ and $t\bar{t} X$ at the LHC.  Our finding is that the DY
production of dimuon basically rules out this class of models shown in Fig.~\ref{fig:diag}(a).
Then in Sec.~\ref{sec:model2}, we propose  a model that can evade the strong constraint from the DY. 
Here we introduce a new vectorlike down-type singlet quark ($B$) and assume the dimuon 
resonance $X$ is a spin-1 particle $Z'$.   
The relevant diagram is shown in Fig.~\ref{fig:diag}(b).
Assuming that only $Z$-$b$-$B$, $Z'$-$b$-$B$ and $Z'$-$\mu$-$\mu$ couplings are nonzero, 
we can identify  the parameter space compatible with the ALEPH 30 GeV dimuon excess. 
Then we discuss  the LHC phenomenology of $B$ quarks, calculating its production cross 
sections and identifying the final states.  
In Sec.~\ref{sec:model3}, we consider a model with a new $U(1)'$ gauge symmetry with 
the associated gauge boson $Z'$ and the singlet scalar
boson $\phi$ charged under 
$U(1)_Y \times U(1)'$.  From the nonzero value of $Z$-$Z'$-$\phi$ vertex, one can account for
$Z\rightarrow Z' \phi \rightarrow \mu^+ \mu^- b\bar{b}$ without conflict with any known 
experimental constraints, but we need a very large $U(1)'$ charge for the $U(1)'$ breaking 
Higgs field.  
In Sec.~\ref{sec:kin}, we obtain the kinematic distributions for
the muon and jets and compare them with the ALEPH data presented in Ref.~\cite{aleph30}.
Finally we will summarize the results in Sec.~\ref{sec:con}.

\section{Simplified Models--I} 
\label{sec:model1}

For a resolution of the 30 GeV dimuon excess, we introduce a new particle $X$,
which could be one of scalar ($s$), pseudoscalar ($a$), vector ($V$),
or axial~vector ($A$).
We assume that the interaction Lagrangian is one of the following forms
for $X=s,a,V,A$:
\begin{subequations}
\begin{eqnarray}
{\cal L}_{\rm scalar} & = & s \sum_f g_f^s \bar{f} f,
\\ 
{\cal L}_{\rm pseudoscalar} & = & i a \sum_f g_f^a \bar{f} \gamma_5 f,
\\
{\cal L}_{\rm vector} & = & - V_\mu \sum_f g_f^V \bar{f} \gamma^\mu f,
\\
{\cal L}_{\rm axial~vector} & = & - A_\mu \sum_f g_f^A \bar{f} \gamma^\mu \gamma_5 f .
\end{eqnarray}
\label{lagrangian}%
\end{subequations}
We consider only two couplings are nonzero: $g_\mu^X$ and $g_b^X$ for $X=s,a,V,A$.
In the pseudoscalar and axial~vector cases, the phenomenology for the $Z$ decay
and LHC phenomenology are similar to the scalar and vector cases, respectively.
Hereafter, we discuss the latter cases unless there is significant change
in the former cases.
For the numerical analysis, we use MadGraph5~\cite{mg5} with implementing the models,  Eqs.~(\ref{lagrangian}).

Then $X$ mainly decays into a $b\bar{b}$ or $\mu^+ \mu^-$ pair.
The decay width of $X$ ($\Gamma^X$) depends on both $g_\mu^X$ and $g_b^X$.
We find that the $X$ boson has a very narrow width for small couplings.
For example,
$\Gamma^V=3\times 10^{-3}~(0.3) $ GeV for $g_\mu^V = g_b^V=0.03~(0.3)$ and
$\Gamma^s=4\times 10^{-3}~(0.4) $ GeV for $g_\mu^V = g_b^V=0.03~(0.3)$,
respectively. In order to achieve a large decay width, $\Gamma^X\sim 1$ GeV,
the couplings should be about $g_f^V\sim 0.6$ or $g_f^s\sim 0.5$, respectively.
If there are other decay channels, the decay width could be enhanced.
However if the decay channels to light quarks or $e^+ e^-$ are open,
much more stringent constraints on the model would be encountered.
Thus one must consider smaller couplings to the light quarks and electron,
which mean that the $X$ must be flavor-dependent. Another possibility
would be the $X$ decay into extra fermions or dark matter candidates,
which often exist in UV complete models with flavor-dependent gauge interactions~\cite{chiralmodel}.
The extra decay channels would be model-dependent and
the UV completion of the model is out of scope of this paper.
In this section, we shall assume that  the $X\to b\bar{b}, \mu^+\mu^-$ decay are the only 
possible decay channels for simplicity.

\begin{figure}[t] 
\includegraphics[width=0.45\textwidth]{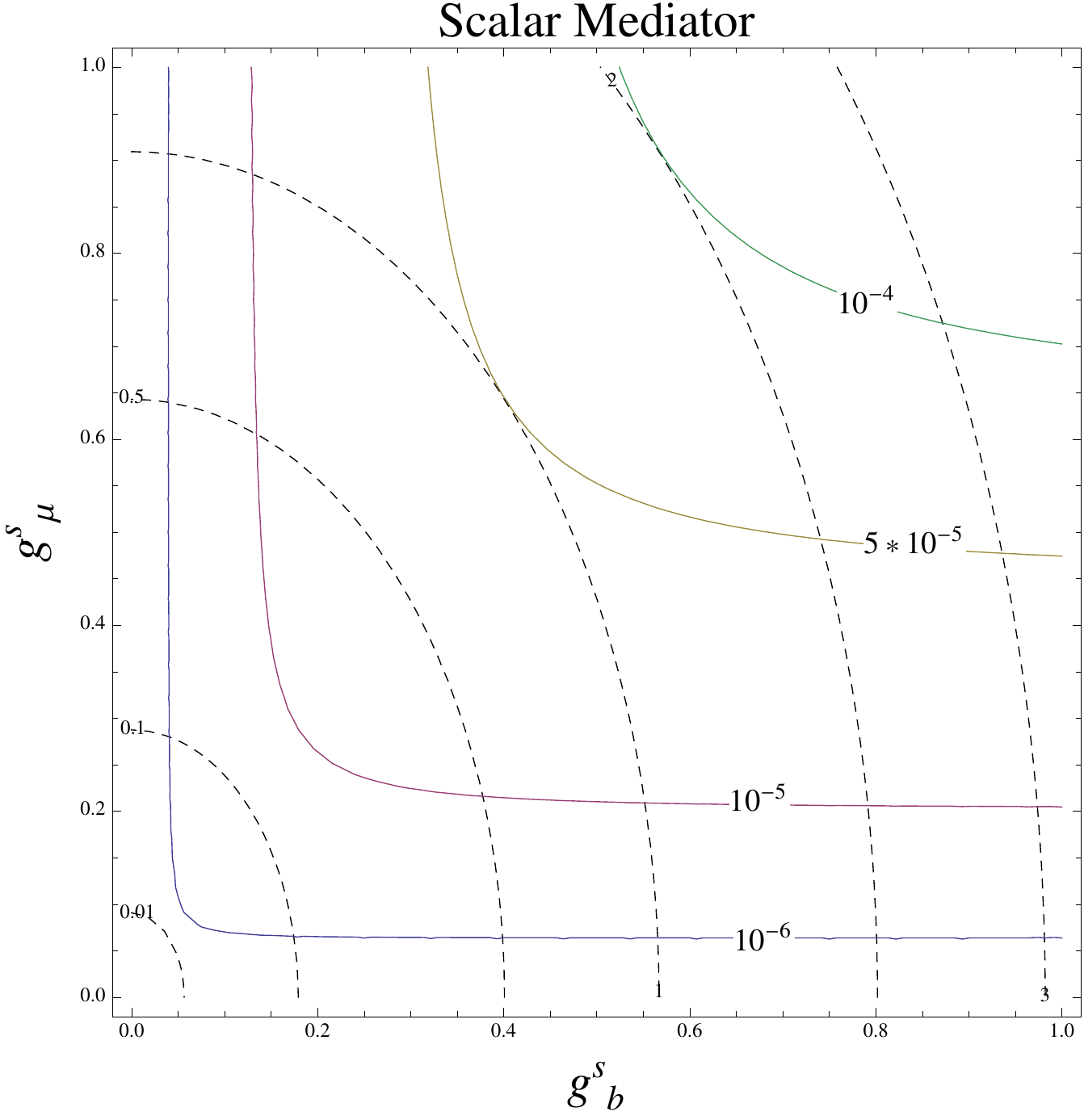}
\includegraphics[width=0.45\textwidth]{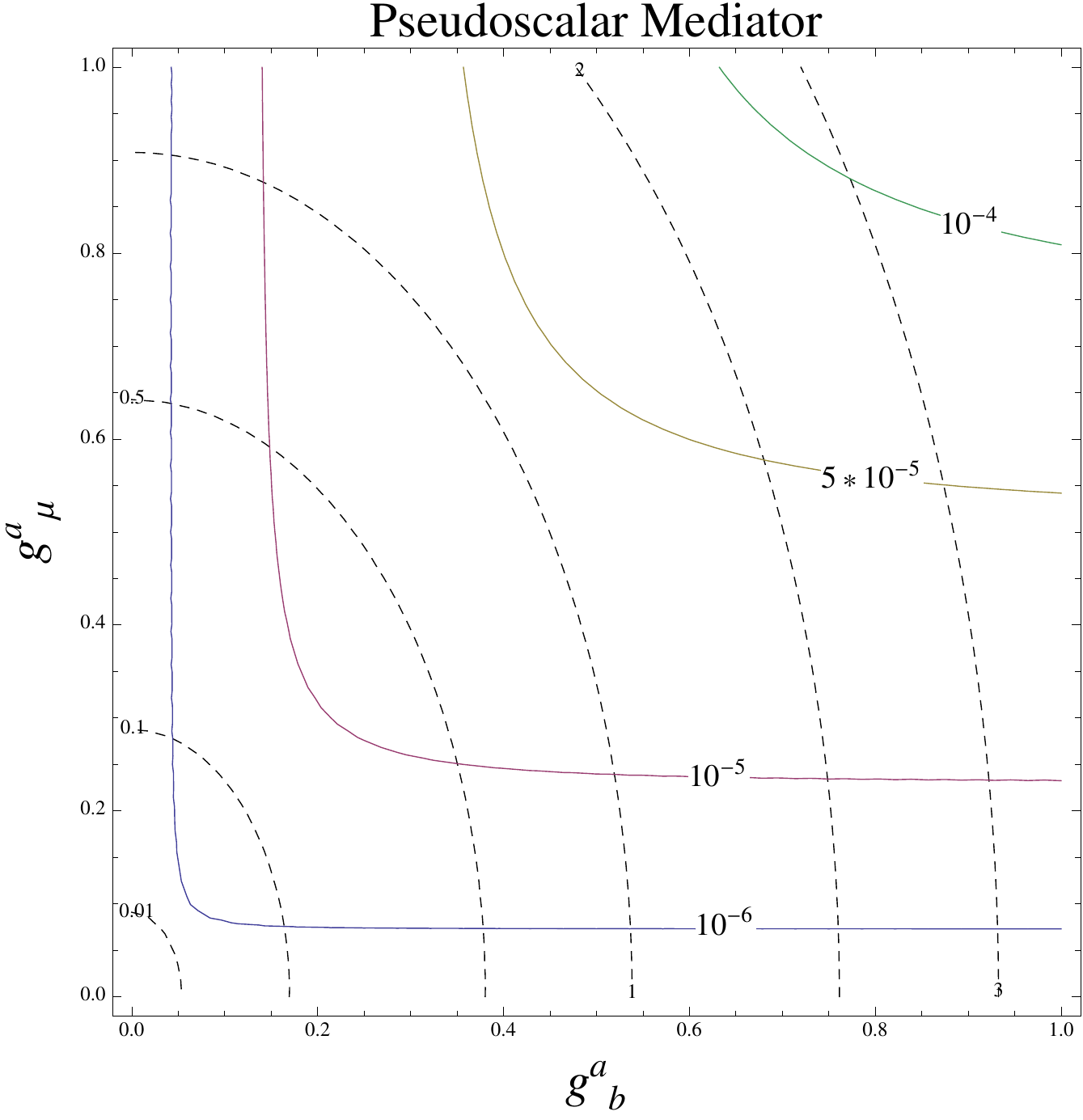}\\
\includegraphics[width=0.45\textwidth]{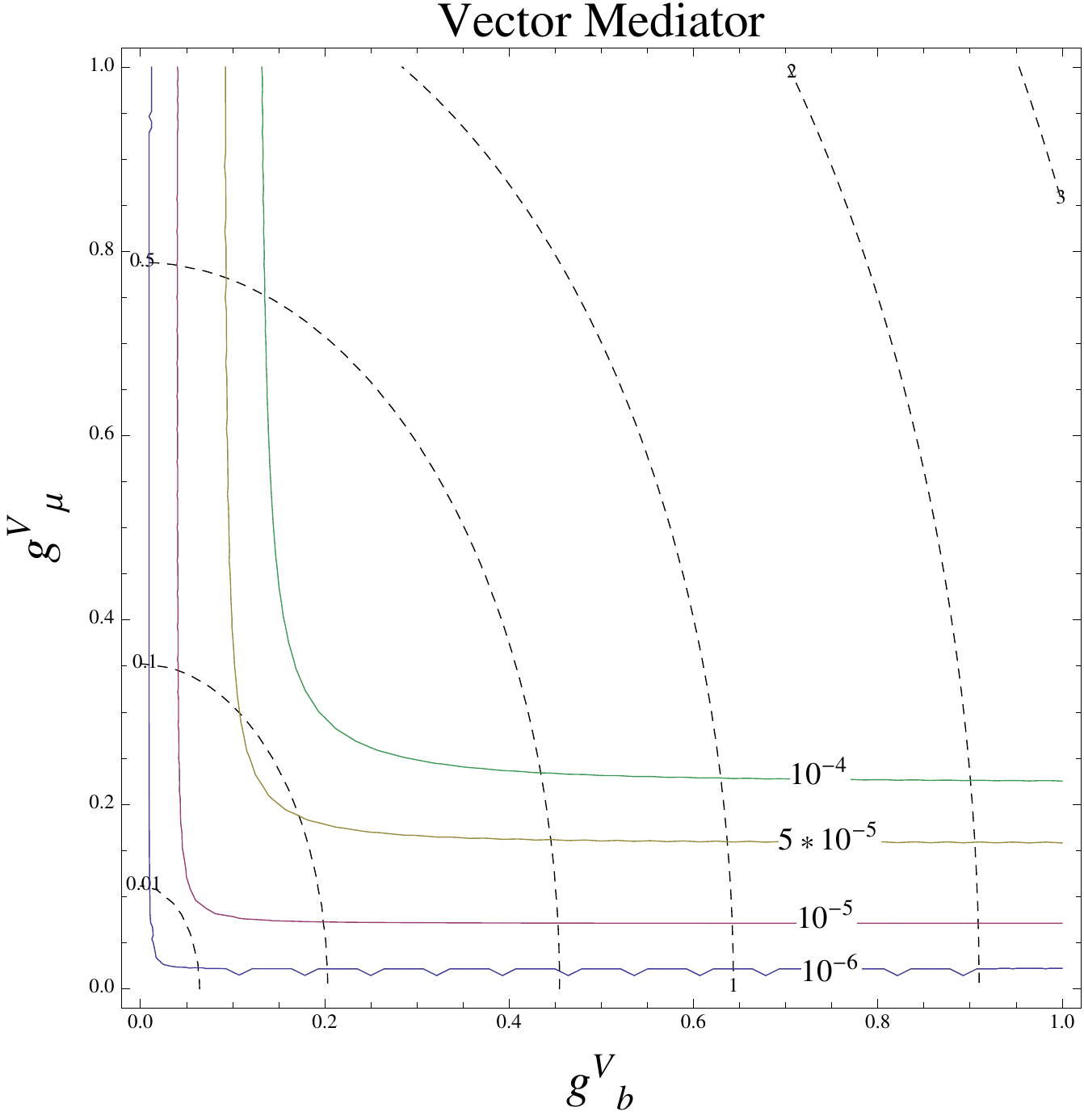}
\includegraphics[width=0.45\textwidth]{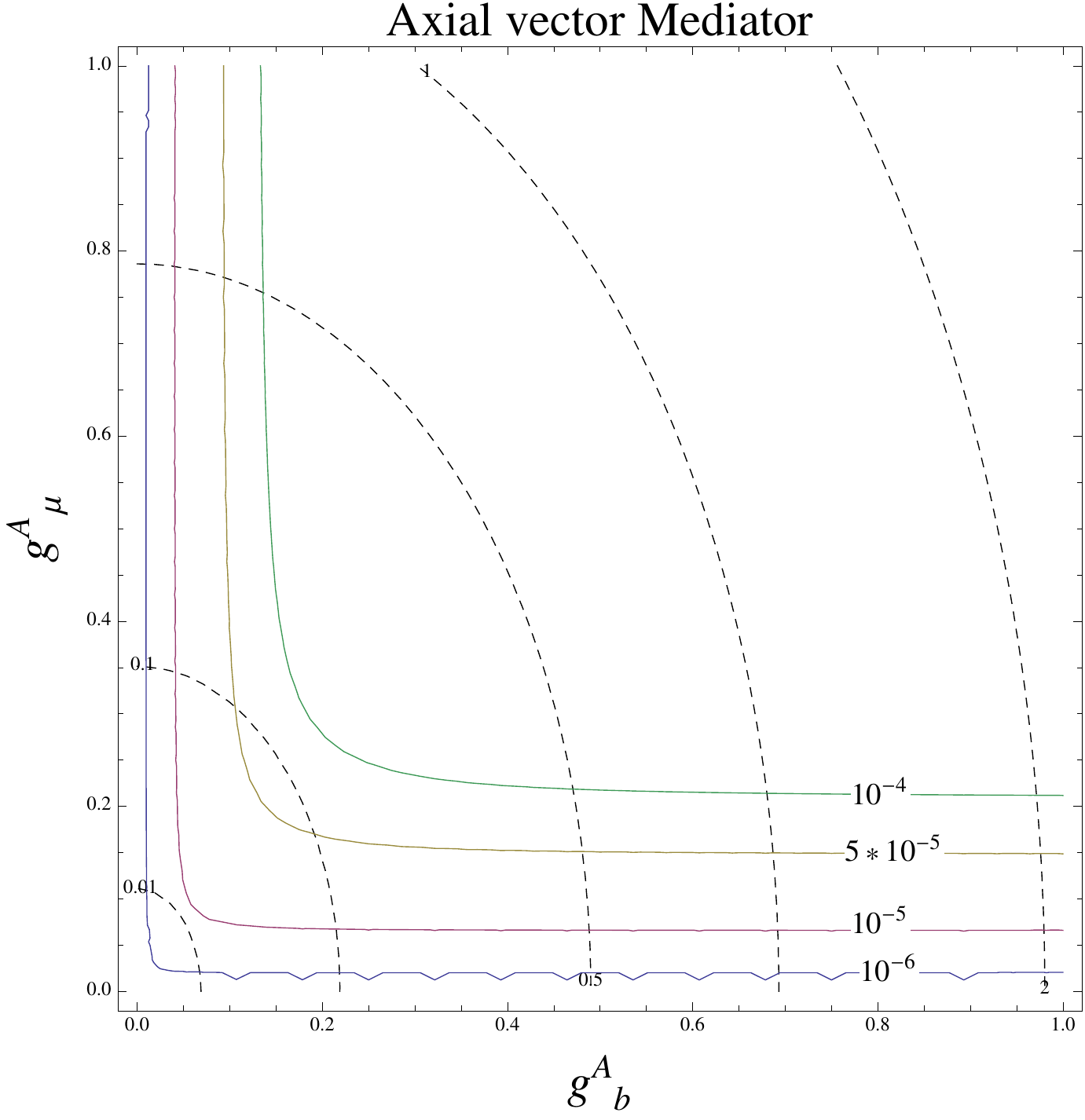}
\caption{\label{fig:Gamx} 
Contour plots for $g_b^X$ and $g_\mu^X$.
The solid lines correspond to $\Gamma^{X}(Z \to b\bar{b}\mu^+\mu^-)$
in unit of GeV
while the dashed lines to the sum of $\Gamma(X\to b\bar{b})$ and
$\Gamma(X\to \mu^+\mu^-)$ in unit of GeV.}
\end{figure}

Since the $X$ boson couples to $b$ and $\mu$, the couplings, $g_b^X$ and 
$g_\mu^X$ can be constrained by the $Z\to 4 b$ and $Z\to 4 \mu$ decays, respectively.
The branching ratio of the $Z\to 4b$ decay is $(3.6\pm 1.3)\times10^{-4}$~\cite{pdg}, whose uncertainty corresponds to $\Delta_{4b}=3.3\times 10^{-4}$ GeV.
We find that the enhancement of the decay width for the $Z\to 4b$ is 
less than $3\times 10^{-4}$ GeV for $g_b^V\lesssim 0.5$ and $g_b^s\lesssim 0.7$,
respectively, which imply that a little bit large couplings are not excluded
by the $Z\to 4b$ decay.
However, the $Z\to 4\mu$ decay might constrain this model significantly.
The branching ratio of the $Z\to 4\ell$ ($\ell=$ either $e$ or $\mu$)
decay was reported by
the CMS and ATLAS collaborations at $\sqrt{s}=7$, $8$, and $13$ TeV~\cite{4mu}.
By the recasting the ATLAS search, the bound on $g_\mu^V$ is obtained
as $g_\mu^V \lesssim 0.025 \sim 0.03$ for $m_V = 30$ GeV
in the $L_\mu-L_\tau$ gauge model~\cite{LmuLtau}.
However, the bound strongly depends on the model, in particular, the total
decay width of $X$.
For more exact bound, the detailed analysis which would depend on the complete 
model and information on cuts in experiments is required.

In Fig.~\ref{fig:Gamx}, we present the contour plots for 
the couplings, $g_b^X$ and $g_\mu^X$ for $X=s$, $a$, $V$, and $A$, respectively.
The solid lines represent
the decay width of the $Z\to b\bar{b}X$ (or $Z\to \mu^+\mu^-X)$ decay 
with subsequent decay $X\to \mu^+\mu^-$  ($X\to b\bar{b}$) in unit of GeV, 
which is denoted by $\Gamma^X(b\bar{b}\mu^+\mu^-)$,
while the dashed lines to the sum of the decay widths of the $X\to b\bar{b}$ 
and $X\to \mu^+\mu^-$, which is approximately equal to the total decay width of the
$X$ boson.
There are other diagrams in the SM,
which interfere with the $X$-mediated diagrams. 
We find that the interference
effects are negligible because the dimuon excess occurs
near the $X$ resonance and its decay width is quite narrow.
The required decay width for the 30 GeV dimuon excess is 
$\Gamma^X(b\bar{b}\mu^+\mu^-)\sim 2.7\times 10^{-5}$ GeV.
In the cases of the scalar and pseudoscalar mediator,
a little large couplings of about $g_\mu^{s,a}\sim g_b^{s,a}\sim 0.4$ 
are required to achieve the dimuon excess at ALEPH.
In this region, the total decay width of the $s~(a)$ 
is about $0.5\sim 0.7$ GeV, which is
marginally consistent with the observed one within the $1\sigma$ level.
We note that it is possible to achieve $\Gamma^{s,a}\sim 1.7$ GeV for
$g_\mu^{s,a}\sim 0.7$ or $g_b^{s,a}\sim 0.7$ while satisfying
the required decay width for $Z\to b\bar{b}\mu^+\mu^-$.
In the cases of the vector and axial vector mediators,
we find that the required decay width for the dimuon excess could be
achieved for $g_\mu^{V,A}\sim g_b^{V,A}\sim 0.015$, but the total decay width
of the $V(A)$ is much smaller; $\Gamma^{V,A}\sim 0.05$ GeV.
We note that $\Gamma^{V,A}\sim 1$ GeV with the required decay width
for the dimuon excess is possible for $g_\mu^{V,A}\sim 0.01$
and $g_b^{V,A}\sim 0.7$, but this region would be excluded by
the $Z\to 4b$ decay.

\begin{table}[t]\centering
 \begin{tabular}{c|c|c|c}
  $g^{Z'}_b$  & $g^{Z'}_\mu$ & $\Gamma^{Z'}(Z\to bb\mu\mu)$  &   $\Gamma(Z' \to bb,\mu\mu)$  \\ 
   0.1 & 0.1 & $2.72 \times 10^{-5}$ & 0.0322  \\ \hline
    $\sigma^{13}(\mu\mu)/\sigma^{1.96}(\mu\mu)$ & $\Gamma(t \to b W Z')$ & $\sigma(pp \to b b Z' )$ & Br$(Z' \to \mu\mu)$  \\
  {\color{red}714.5}/{\color{red}55.8} pb & $1.267 \times 10^{-4}$ & 136.1 pb & 0.25 \\
 \end{tabular}
 \caption{Benchmark point I (vector mediator) at 13 TeV LHC and Tevatron. }
\label{table1}
\end{table}

\begin{table}[t]\centering
 \begin{tabular}{c|c|c|c}
  $g^{Z'}_b$  & $g^{Z'}_\mu$ & $\Gamma^{Z'}(Z\to bb\mu\mu)$  &   $\Gamma(Z' \to bb,\mu\mu)$  \\ 
   0.7 & 0.1 & $3.036\times 10^{-5}$  & 1.19   \\ \hline
    $\sigma^{13}(\mu\mu)/\sigma^{1.96}(\mu\mu)$ & $\Gamma(t \to b W Z')$ & $\sigma(pp \to b b Z' )$ & Br$(Z' \to \mu\mu)$  \\
  {\color{red}920.5}/{\color{red}71.1} pb & 0.0062 & 6645 pb & 0.0068 \\
 \end{tabular}
 \caption{Benchmark point II (vector mediator) at 13 TeV LHC and Tevatron. }
\label{table2}
\end{table}

Next, we consider some phenomenology of the vector-mediator model at the LHC
with $\sqrt{s}=13$ TeV (and at the Tevatron).
Now we call the $V$ as a $Z^\prime$ boson.
Since the $Z^\prime$ couples with $b$ quark and $\mu$,
the processes which include both or either of $b$ or $\mu$ would provide
stringent constraints. 
Among them, we investigate three processes: the Drell-Yan process,
$pp\to Z^\prime \to \mu^+\mu^-$, the top decay, $t\to b W Z^\prime$,
and the $Z^\prime$ production associated with a $b\bar{b}$ pair,
$pp\to b\bar{b}Z^\prime$.

Here, we consider two benchmark points.
In Table~\ref{table1}, the couplings are taken to be $g_\mu^{Z'}=g_b^{Z'}=0.1$,
which yield $\Gamma^{Z'}(Z\to bb\mu\mu)=2.72\times 10^{-5}$ GeV.
Since the branching ratio of $Z^\prime \to \mu\mu$ is $0.25$,
the cross section for $pp\to bbZ^\prime$ with the subsequent decay 
$Z^\prime\to \mu\mu$ would be about 34 pb, which can be a probe of this model. 
For the top decay,
the branching ratio of $t \to b W Z'$ followed by $Z' \to \mu\mu$
would be about $2\times 10^{-5}$, which could be another probe at the LHC.
The most constraining process for this class of models is the Drell-Yan process, 
whose cross section is about $715$ pb at the LHC and about $55.8$ pb at the Tevatron, 
which is already excluded by the CMS data~\cite{dy}.

In Table~\ref{table2}, we take the couplings to be
$g_\mu^{Z'}=0.1$ and $g_b^{Z'}=0.7$,
which yield $\Gamma^{Z'}(Z\to bb\mu\mu)=3.04\times 10^{-5}$ GeV.
As we already discussed, this set would be excluded because it predicts
a large decay width for $Z\to 4b$ via the $Z'$ exchange.
Actually, the decay width is about $1.7\times 10^{-3}$ GeV, which is much
larger than the PDG value.
The cross section for $pp\to bbZ'$ is about $6600$ pb and, then,
it for $pp\to bbZ'$ followed by $Z'\to \mu\mu$ is about 45 pb, which can be
easily proved at the LHC. In this case, the Drell-Yan process strongly 
constrains this model, too. 
The predicted cross section reaches about $900$ pb
at the LHC and about $71.1$ pb at the Tevatron,
which is already excluded by the CMS data~\cite{dy}. 

We also investigate some benchmark points in the scalar-mediator model.
We find that the cross section for the Drell-Yan cross section 
exceed 380 pb at $\sqrt{s}=13$ TeV for $g_b^s=g_\mu^s=0.1$, which is also
excluded by the CMS data.
We note that those couplings are too small to provide the required decay width for the dimuon excess.

Since the $X$ boson can couple to both $b\bar{b}$ and $\mu^+ \mu^-$
in the models discussed in this section,
one may observe a similar peak of the $X$ boson for the invariant mass
of the $bb$ pair, depending on the couplings of the $X$ . 
So far the $b\bar{b}$ peak in $Z\rightarrow b\bar{b} \mu^+ \mu^-$ has not been observed 
in the $Z$ decay yet.  The search for the resonance of the $b\bar{b}$ pair would be another  
probe of those models.

Summarizing this section, we investigated some benchmark points to predict the required
decay width for the ALEPH 30 GeV dimuon excess reported in Ref.~\cite{aleph30}
in simplified models defined by Eqs.~(\ref{lagrangian}a)--(\ref{lagrangian}d) and Fig.~\ref{fig:diag}(a).
In general, one can find the preferred points, but they predict too large 
cross sections or branching ratios in other productions or decay channels.
Most notably the Drell-Yan process turns out to be the most stringent constraints, 
and the simplest benchmark points (and also other points) are easily excluded
by the LHC data, in particular, by the Drell-Yan process.
One might decrease the Drell-Yan cross section by increasing
the decay width of $X$, but it necessarily decreases the decay width
for $Z\to bb\mu\mu$ and the dimuon excess would not be explained with
the $X$ boson with interactions defined in Eqs.~(\ref{lagrangian}a)--(\ref{lagrangian}d).

\section{Simplified Model--II} 
\label{sec:model2}

The Model--I in Sec.~\ref{sec:model1} has a problem with DY production through $b\bar{b} \rightarrow Z' 
\rightarrow \mu^+ \mu^-$.  One can avoid this problem by making $Z'$ decouple from 
$b \bar{b}$ and introducing a new vectorlike down-type quark $B$ which has a nonzero 
$Z$-$b$-$B$ and $Z'$-$b$-$B$ couplings.

\begin{figure}[t] 
\includegraphics[width=0.45\textwidth]{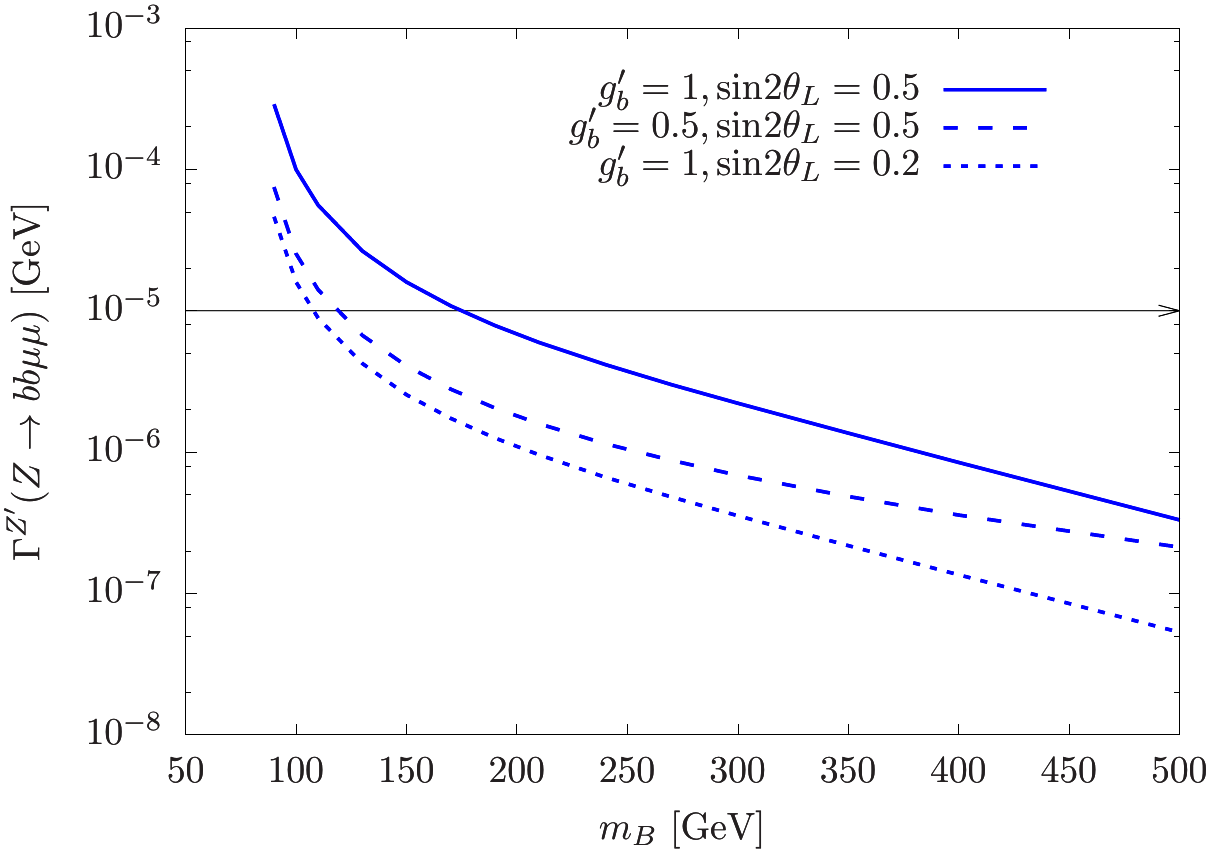}
\includegraphics[width=0.5\textwidth]{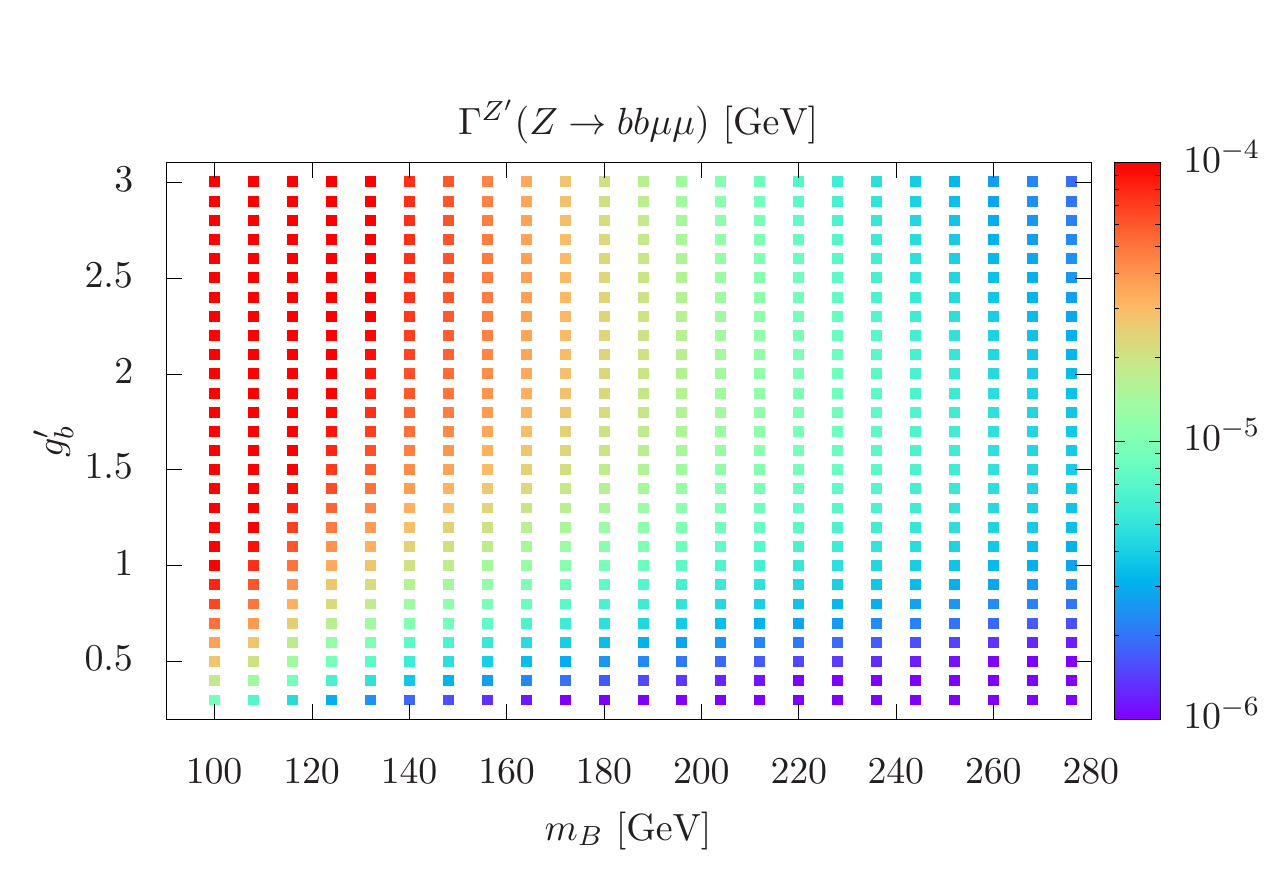}
 \caption{\label{fig:vecxsec} The $\bar{b}b\mu\mu$ partial decay width of the Z boson from new physics contribution. $\sin\theta_L=0.5$ in the right panel.}
\end{figure}

Let us assume a down-type vectorlike singlet quark $B$ with the following interactions (see Ref.~\cite{Aguilar-Saavedra:2013qpa}):
\begin{align}
\mathcal{L}= g'_\mu Z'_\rho \bar{\mu}  \gamma^\rho \mu  
+ g_s G^a_\mu \bar{B} \gamma^\mu T^a B - 
 \left[ \frac{1}{2} g'_b Z'_\rho \bar{b}\gamma^{\rho}B  + \frac{g_W \sin 2\theta_L}{4 c_W} Z_\mu \bar{b}\gamma^\mu P_L B + h.c.\right] .
\end{align}
Here the $Z'$ coupling to the quark sector is assumed  to  always involve the vectorlike quarks 
so that $Z' \rightarrow b \bar{b}$ is zero (or highly suppressed) and 
the DY process is not allowed (or highly suppressed). 

The ALEPH 30 GeV dimuon excess in the $Z$ decay is explained by  
$Z \rightarrow b B^* \rightarrow bb \mu^+ \mu^-$ assuming that $B$ is heavier than $Z$ so that $Z \rightarrow B \bar{b} + C.C.$ involves a virtual $B$ quark.  The main decay channels of $B$ would be  
$B \rightarrow b Z'$. As for $Z'$,  we assume that the $Z' \rightarrow \mu^+ \mu^-$  is the only 
kinematically  allowed decay channel, so the $g'_\mu$ parameter is not relevant to the partial decay
width of $Z \rightarrow b B^* \rightarrow bb \mu^+ \mu^-$ decay and will be set to 
$g'_\mu=0.01$ according to the $Z\to4\mu$ measurement~\cite{LmuLtau}. 

In Fig.~\ref{fig:vecxsec}, we show the partial width of $Z \rightarrow b B^* \rightarrow bb 
\mu^+ \mu^-$ ($\Gamma^{Z'} (Z \rightarrow bb\mu\mu)$) in terms of the parameters of the 
Model--II, i.e. $m_B$, $g'_b$ and $\sin 2 \theta_L$. From the left panel of Fig~\ref{fig:vecxsec}, we can observe 
the $\sin^2 2 \theta_L$ behavior of $\Gamma^{Z'} (Z \rightarrow bb\mu\mu)$, while the 
dependence on $g'_b$ is more complicate because it could change the total width of $B$ 
which is important for $\Gamma^{Z'} (Z\to bb\mu\mu)$. So in the right panel, fixing 
$\sin\theta_L=0.5$,  we show the $\Gamma^{Z'} (Z\to bb\mu\mu)$ in the 2-dimensional plane of 
$g'_b$ and $m_B$. From the figure, we can find the ALEPH dimuon excess can be addressed 
in the parameter space with $m_B\sim 100-200$ GeV, $g'_b \sim 0.5-3$ and 
$\sin\theta_L \sim 0.5$, requiring that $g'_b$ remains within the perturbative regime.

\begin{table}[t]\centering
 \begin{tabular}{c|c|c|c}
  $g'_b$  & $m_B$ & $\Gamma^{Z'}(Z \to bb\mu\mu)$  &   $\Gamma(B\to b Z')$  \\ 
   0.7 & 110 GeV & $2.75 \times 10^{-5}$  &   3.4 GeV   \\ \hline 
 Br$(B\to b Z')$  &  $\sigma^{13}(BB)/\sigma^{8}(BB)$ & $\sigma^{13}(bB)/\sigma^{8}(bB)$ & $\sigma^{13}(qB)/\sigma^{8}(qB)$  \\
 1.0  & 3942/1203 pb  & 1.68/0.89 pb & 7.49/3.2 pb \\ 
 \end{tabular}
 \caption{\label{tab:bpiii} Benchmark point III (vectorlike particle model). }
\end{table}

\begin{table}[t]\centering
 \begin{tabular}{c|c|c|c}
  $g'_b$  & $m_B$ & $\Gamma^{Z'}(Z \to bb\mu\mu)$  &   $\Gamma(B\to b Z')$  \\ 
   2.0  & 180  & $2.35\times 10^{-5}$ GeV &  124.4 GeV  \\ \hline  
 Br$(B\to b Z')$  &  $\sigma^{13}(BB)/\sigma^{8}(BB)$ & $\sigma^{13}(bB)/\sigma^{8}(bB)$ & $\sigma^{13}(qB)/\sigma^{8}(qB)$  \\
 1.0  &  391/120 pb & 0.21/0.1 pb & 4.24/1.67 pb \\ 
 \end{tabular}
 \caption{\label{tab:bpiv} Benchmark point IV (vectorlike particle model). }
\end{table}

Two benchmark points with appropriate dimuon excess and different masses 
for the vectorlike 
$B$ are given in Table~\ref{tab:bpiii} and Table~\ref{tab:bpiv}. 
For the relatively heavy $B$, a coupling $g'_b$ that approaches 
the perturbativity limit ($g'_b=2$) is required, 
rendering very large decay width of 
$B$ whose main decay channel is assumed to be $b \mu \mu$. 
It will be important to study the signals of those two benchmark points at the LHC. 
The process with the largest cross section is the pair production of $B\bar{B}$ through QCD 
coupling, followed by the $B\to b Z'$ decay which produces the $bb+ 4\mu$ signal. The corresponding production cross section at the 13/8 TeV LHC for benchmark points ($\sigma^{13}(BB)/\sigma^{8}(BB)$) are given in the Tables III and IV for $m_B = 110$ GeV and 180 GeV, 
respectively. The cross section is quite sensitive to the mass of $B$, raising the $m_B$ from 110 GeV to 180 GeV can reduce the cross section by one order of magnitude. 
Even though there is no LHC search for $bb + 4\mu$ final states so far, the signal is almost background free and 
$b\bar{b}$ + multi muon events are rare. Therefore it is very likely that  this kind of model has been excluded 
already \cite{private}. 

Because of the $b$-$B$-$Z$ coupling, there could also be signals of $p p \to Z \to b B$ 
and $q b\to q B$ with $t-$channel $Z$ exchange, where $q=u,d,s,c$. These processes will 
generate the signal of $b\mu\mu$ that is associated with an additional $b$-jet or light flavor jet, 
the cross sections 
of which are proportional to $\sin^2 2 \theta_L$. Those cross section for benchmark points are 
also presented in Table~\ref{tab:bpiii} and Table~\ref{tab:bpiv}, where we find both processes have production cross section of $\mathcal{O}(1)$ pb at the 13/8 TeV LHC and t-channel process has much larger production rate than the s-channel process. Moreover, as we can expect, the production cross section of the $t-$channel process is less sensitive to $m_B$ than that of the $s-$channel process.  It would be highly desirable to look into  $bb+ 4\mu$, 
$bb + 2\mu$, and $bj+2\mu$ more carefully at the LHC and test this model for the ALEPH 30 GeV dimuon excess. 

\section{New gauge $U(1)'$ model} 
\label{sec:model3}

In this section, we introduce a model with new $U(1)'$ gauge symmetry (Model--III), 
which includes a new gauge boson $Z'$ and a new real scalar $\phi$ 
(see Fig.~\ref{fig:diag}(c)). 
Here, we do not focus on the details of this model, but we just mention that the $U(1)'$ could 
be the $U(1)_{L_\mu-L_\tau}$ where $Z' \rightarrow e^+ e^-$ can be naturally suppressed. 
The scalar $\phi$ is assumed to be charged under both $U(1)_Y$ and $U(1)'$ gauge groups, 
with their charges being $Y_\phi$ and $Y'_\phi$, respectively.
The $U(1)'$ is spontaneously broken by the nonzero VEV of $\phi$, $v_\phi$. 
Since $\phi$ is charged under  $U(1)_Y \times U(1)'$ gauge symmetries, there appear  three neutral spin-1 
fields,  $B_\mu$, $W^3_\mu$, and   $\hat{Z}'_\mu$, which  are mixed with each other after electroweak and 
$U(1)'$  symmetry breaking.  Then one can obtain three mass eigenstates, $A_\mu$ (massless photon), 
$Z_\mu$, and  $Z'_\mu$.   Given that the covariant derivative for $\phi$ is written as 
\[
D_\mu \phi = \partial_\mu \phi - i g_1 Y_\phi B_\mu \phi -i g' Y'_\phi Z'_\mu \phi , 
\] 
one can write down the mass matrix for three neutral gauge bosons in the interaction eigenstates, 
$(B_\mu, W^3_\mu, \hat{Z}'_\mu)$ as
\begin{equation}
M_V^2 = 
\left(
\begin{array}{ccc}
 g_1^2 \frac{v^2}{8}  + g_1^2 Y^2_\phi v^2_\phi 
& -g g_1 \frac{v^2}{8} 
& g_1 g' Y_\phi Y'_\phi v^2_\phi \\
 -g g_1 \frac{v^2}{8} & g^2 \frac{v^2}{8} & 0 \\
 g_1 g' Y_\phi Y'_\phi v^2_\phi & 0  
& g'^2 Y_\phi^{\prime 2}  v^2_\phi 
\end{array}
\right),
\end{equation}
where $g_1$, $g$, and $g'$ are the couplings of the $U(1)_Y$, $SU(2)_L$,
and $U(1)'$, respectively.

We shall simply assume the $U(1)'$ gauge symmetry is related to the $\mu$--flavor so that $Z'$ 
can dominantly decay into $\mu^+ \mu^-$.  On the other hand the scalar $\phi$ would decay mainly  
into $b\bar{b}$  through its mixing with the SM Higgs boson, with the Yukawa coupling given by 
\[
\frac{m_b}{v} \sin\alpha  \sim \frac{1}{50} \sin\alpha  \lesssim 10^{-3} .
\]   
We have used the fact that the current bound on the singlet-Higgs mixing angle ($\alpha$) from the 125 GeV Higgs
signal strengths is  $\sin\alpha \lesssim 0.4$ depending on the $\phi$ mass \cite{Cheung:2015dta,Dupuis:2016fda}.

\begin{figure}[t!] 
\includegraphics[width=0.45\textwidth]{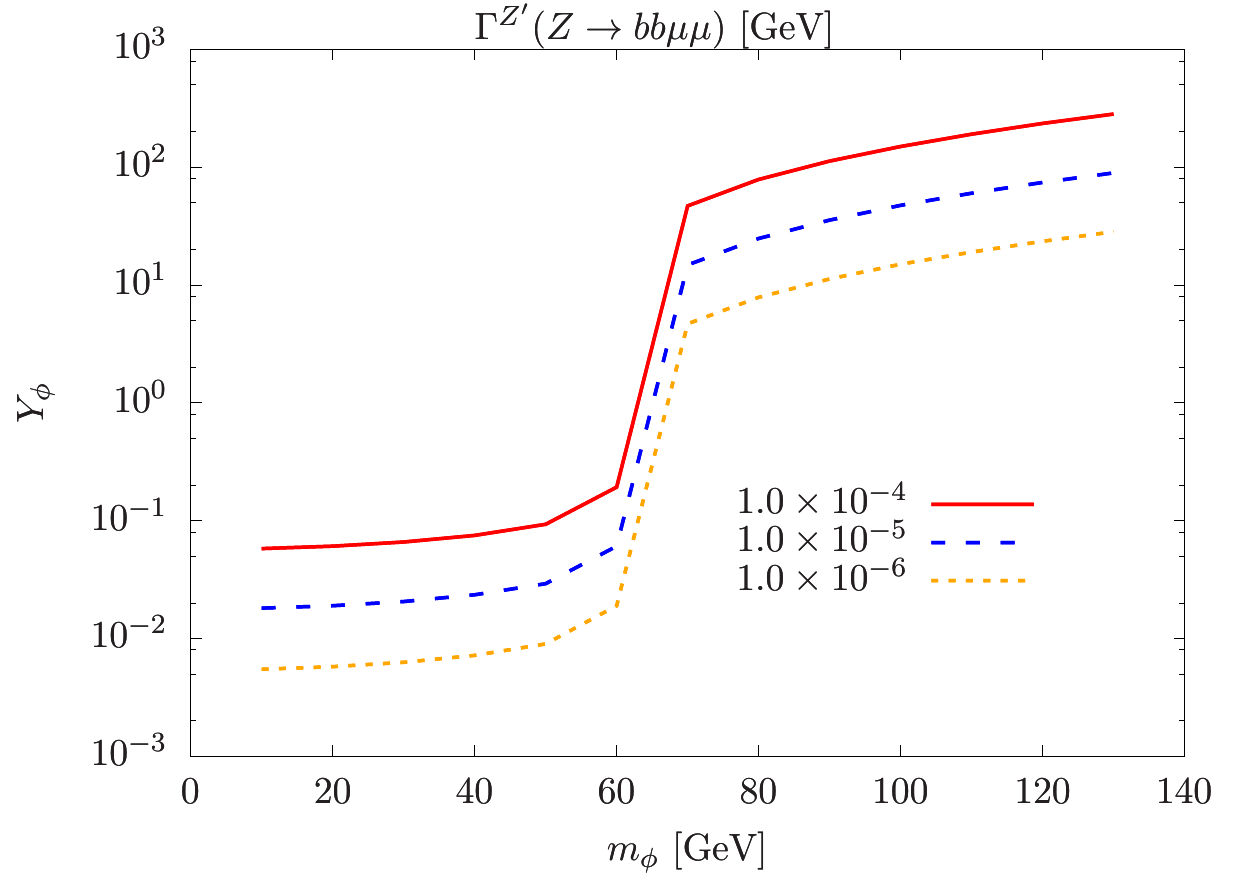}
 \caption{\label{fig:u1xsec} The contours  of $\Gamma^{Z^{\prime}}(Z \to bb \mu\mu)$ with varying $m_\phi$ and $Y_\phi$. The dimuon excess requires $\Gamma^{Z^{\prime}}(Z \to bb \mu\mu) \sim 2\times 10^{-5}$ GeV. }
\end{figure}

The following constraints have to be applied to be consistent with experimental measurements 
in order that we try to explain the dimuon excess: 
\begin{itemize}
\item The mixing between $Z'$ and SM gauge field should be small, 
so we have $(M_V^2)_{33} \sim \frac{1}{2} m^2_{Z'}$ where $m_{Z'} \sim 30$ GeV 
according to the observed excess. 
Then, we can express $v_\phi = \frac{m_{Z'}}{\sqrt{2} g' Y'_\phi }$.
\item The current measurements imply the mixing between $Z'$ and SM $Z$ 
to be $\sin \theta_{Z Z'} \lesssim \mathcal{O}(10^{-2})
\sim \mathcal{O}(10^{-3})$~\cite{zzpmixing}
and the mixing between $Z'$ and photon to be 
$\sin \theta_{\gamma Z'} \lesssim \mathcal{O}(10^{-2})$~\cite{photonzpmixing}
for $m_{Z'} = 30$ GeV. 
This requires very small off-diagonal component of the gauge boson mass matrix, 
e.g. $(M_V^2)_{13}< 5$ GeV$^2$, which corresponds to $\sin \theta_{Z Z'}< 5.1 \times 10^{-4}$ and $\sin \theta_{\gamma Z'}<9.8\times 10^{-3}$ result in $\frac{Y_\phi}{Y'_\phi}\lesssim 3.2 \times 10^{-2} g'$. 
\item As we have discussed in Sec.~\ref{sec:model1}, the $g'$ should be $g'\lesssim 0.02$ in order to suppress the $Z\to 4\mu$ decay. 
Then we find the condition for the ratio of $U(1)$ charges, 
$\frac{Y'_\phi}{Y_\phi} \gtrsim 1575$, which implies a large gap between
two $U(1)$ charges or $Y_\phi =0$. 
\item The $Z$-$Z'$-$\phi$ coupling is
$C_{ZZ'\phi} = \frac{g_1}{\sqrt{g_1^2 + g^2}} 2 g_1 Y_\phi m_{Z'} + \sin \theta_{ZZ'} \sin \alpha \frac{m_Z^2}{v}$, 
where only $Y_\phi$ is the free parameter. Taking $\sin \theta_{ZZ'}=5.1 \times 10^{-4}$ (correspond to $(M_V^2)_{13}=5$ GeV$^2$) and $\sin \alpha=0.2$, we plot the contours 
of $\Gamma(Z\to bb Z')$ on the $m_\phi$-$Y_\phi$ plane 
in Fig.~\ref{fig:u1xsec}. Assuming Br$(Z' \to \mu\mu)=100\%$, $Y_\phi$ 
in very large range of $\mathcal{O}(10^{-2})-\mathcal{O}(10^2)$ can explain the dimuon excess.
\item Finally $\phi \rightarrow b  \bar{b}$ will occur through $\phi-h$ mixing, whose mixing angle
$\sin\alpha$ is constrained to be $\sin\alpha \lesssim 0.2 $ by the present LHC data on the
Higgs signal strengths. Including the Yukawa coupling of the SM Higgs to the SM fermions, 
$\phi$-$b$-$\bar{b}$ will be given by $( m_b /v ) \sin\alpha \sim (1/50) \sin\alpha$. 
\end{itemize}

In this model, we find that a very large $U(1)'$ charge for the scalar $\phi$ is required 
in order to accommodate the observed dimuon excess.   For a reasonable hypercharge,
say $Y_\phi \sim 1$, the required value  for $Y'_\phi$ is $\mathcal{O}(10^4)$. 
This implies that a reasonable or natural model building for the $U(1)'$ model
might be difficult to be accommodated with the dimuon excess. 

The mixing of the $Z$ and $Z'$ may also be generated by the kinetic mixing,
$\frac{\varepsilon}{2 c_W} \hat{B}_{\mu\nu} \hat{Z}'^{\mu\nu}$~\cite{Lee:2016ief}. 
Assuming no mass mixing ($Y_\phi=0$) for simplicity, we find that
the coupling related to the $Z$ boson decay is
\begin{align}
C_{ZZ'\phi} = (\sin \xi - \varepsilon t_W \cos \xi) \sqrt{2} g' Y'_\phi m_{Z'} + \sin \xi \sin \alpha \frac{m_Z^2}{v},
\end{align}
with the mixing angle $\tan 2 \xi = \frac{2 \varepsilon t_W}{1- (\varepsilon t_W)^2 - m^2_{Z'} /m^2_{Z}}$.
The model independent bound on kinetic mixing when $m_{Z'} =30$ GeV is $\varepsilon \lesssim 0.03$~\cite{zzpmixing}. In the small $\varepsilon$ limit, 
\begin{align}
\sin \xi &=0.62 \varepsilon  + \mathcal{O}(\varepsilon^3),\\
\cos \xi &= 1- 0.2 \varepsilon^2 + \mathcal{O}(\varepsilon^4).
\end{align}
so we get $C_{ZZ'\phi} \sim (3.04 g' Y'_\phi +20.87 \sin \alpha )\varepsilon$ GeV. 
Taking $\varepsilon =0.03$ and $\sin \alpha =0.2$, we find that $g'Y'_\phi$ should be at least $\sim 1.3$ in order to have $\Gamma(Z\to Z' \phi) \gtrsim 1.8 \times 10^{-5}$ GeV. 
Since $g' \lesssim 0.02$ from $Z\to 4\mu$ constraint, $Y'_\phi \gtrsim 65$ is required. 
Again, we find that a large $Y'_\phi$ is required in order to accommodate
with the ALEPH dimuon excess.

\begin{figure}[t] 
\includegraphics[width=0.45\textwidth]{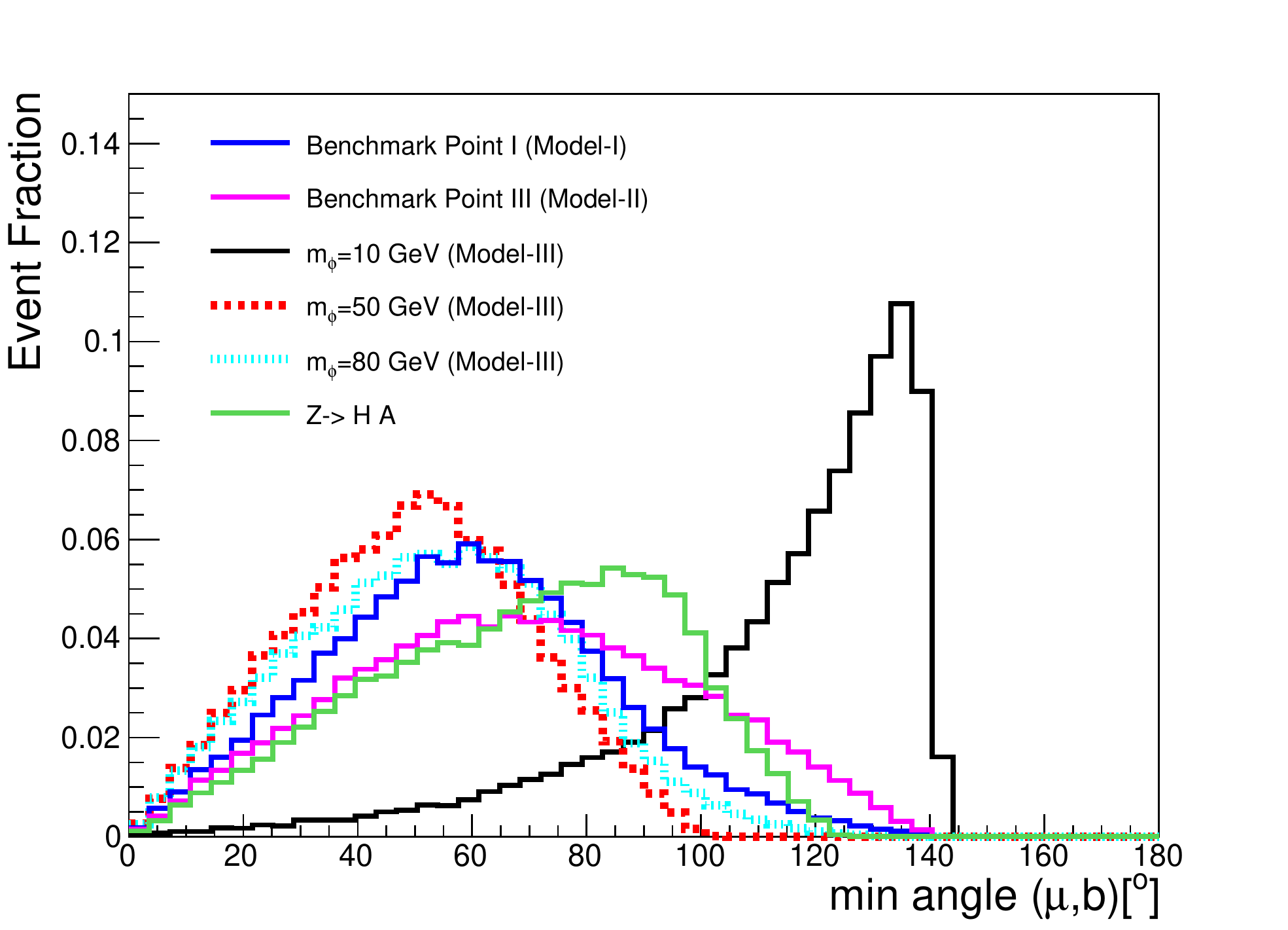}
\includegraphics[width=0.45\textwidth]{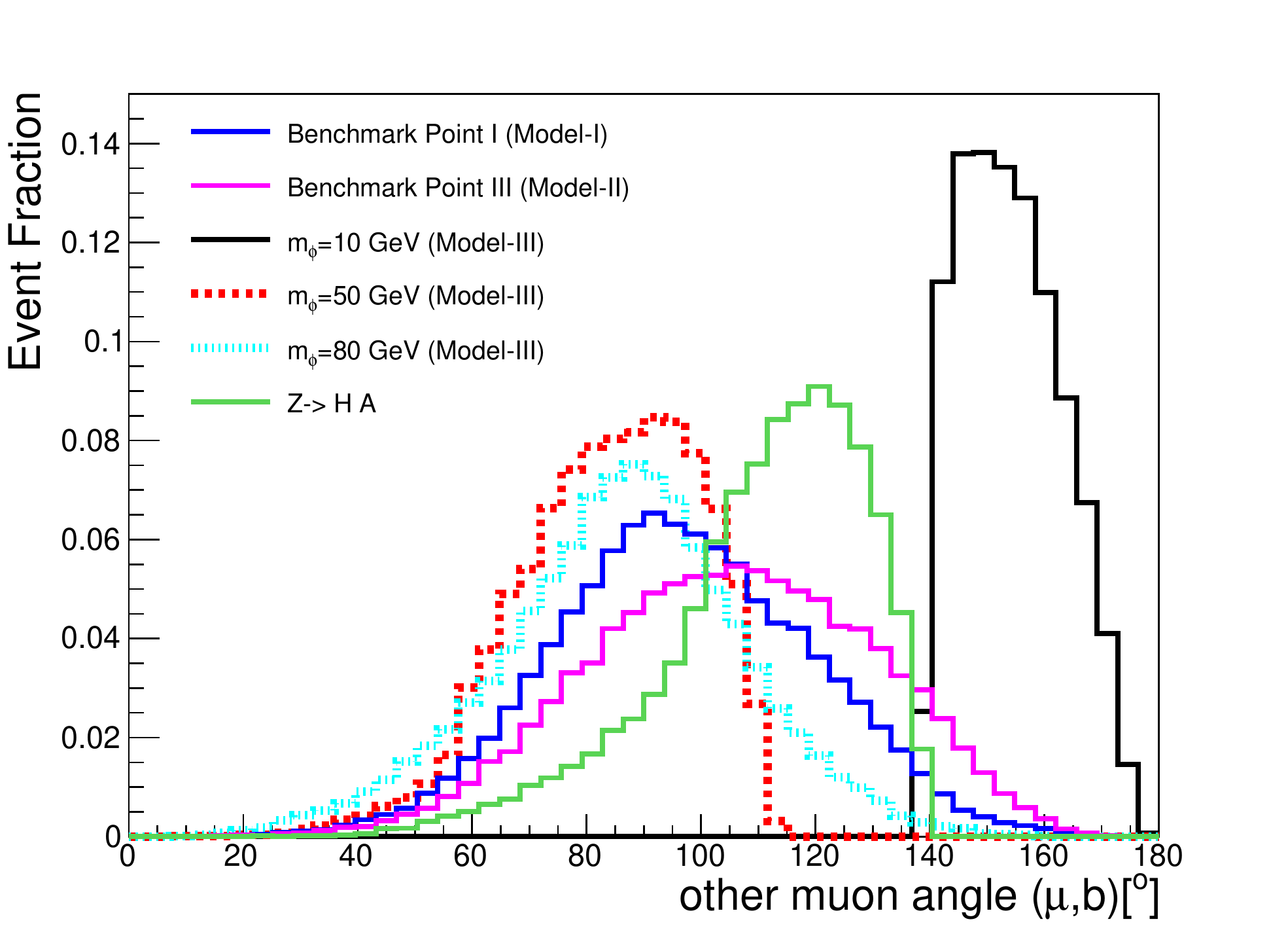}
\includegraphics[width=0.45\textwidth]{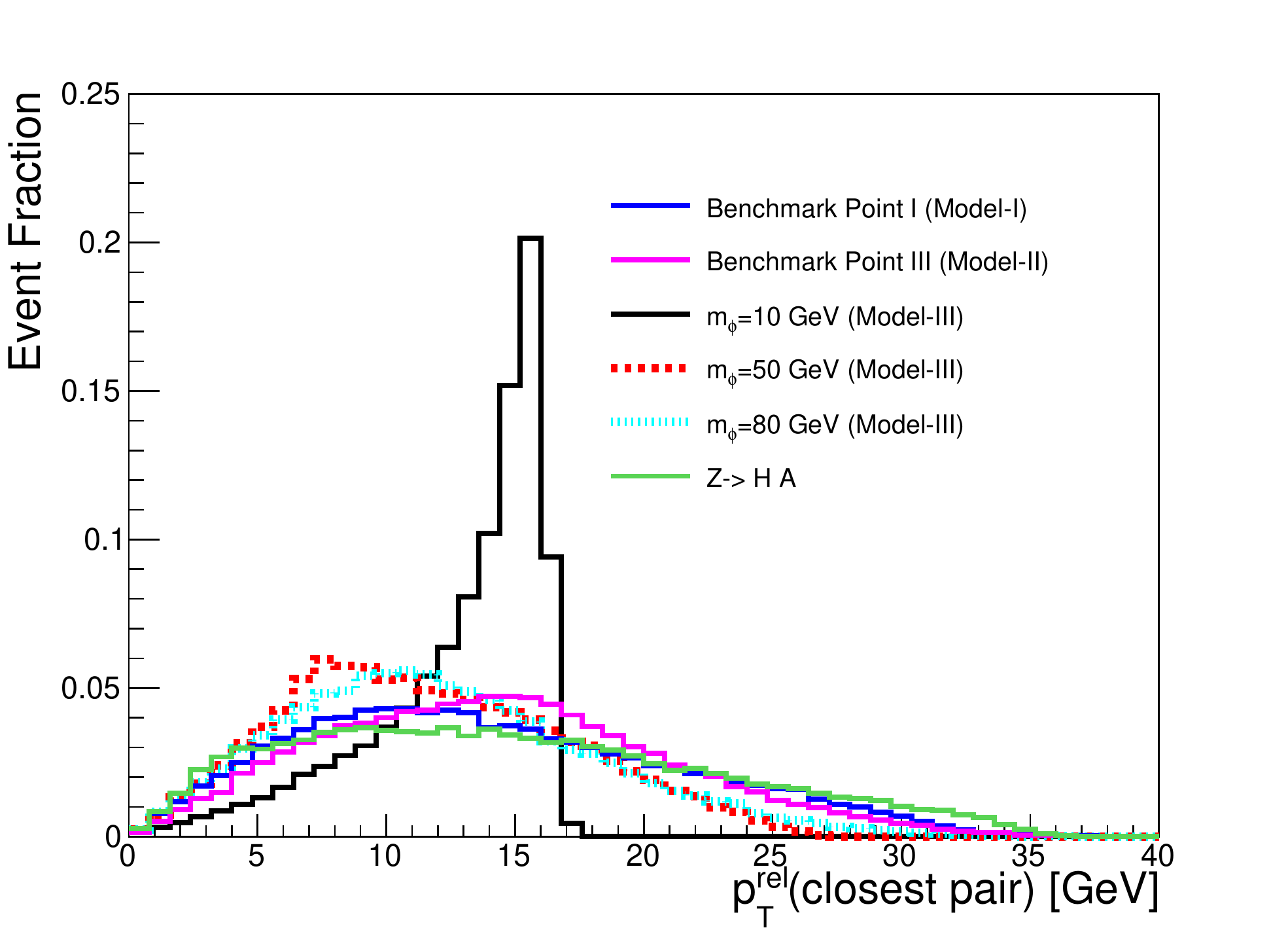}
\includegraphics[width=0.45\textwidth]{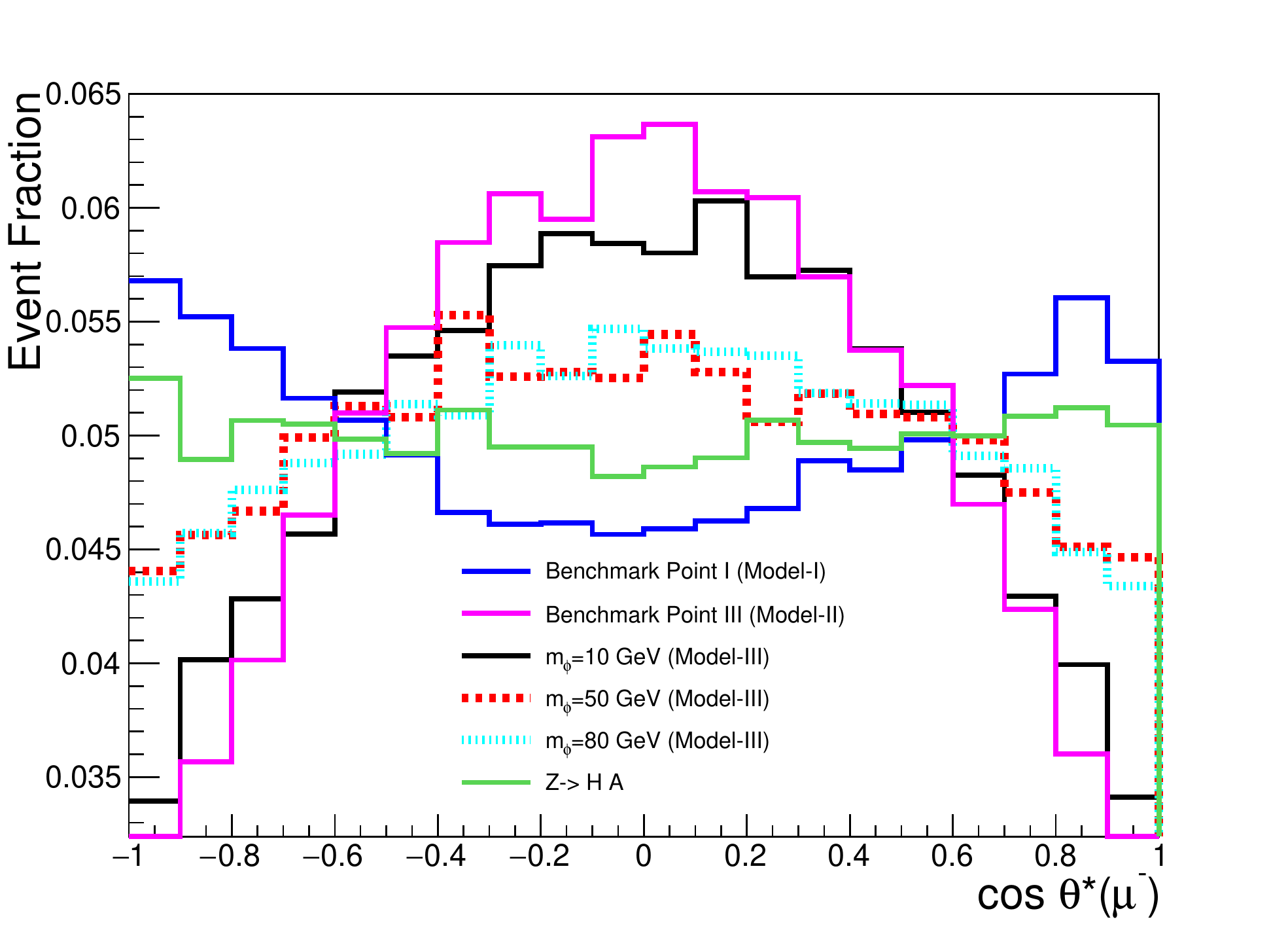}
 \caption{\label{fig:kin} The event fractions as functions of
the minimum angle between a muon and a leading jet (top left panel), 
the other muon angle defined in the text (top right panel), 
the relative transverse momentum of the closest muon-jet pair (bottom left panel), 
and 
the decay angle $\cos\theta^*$ distribution for muons ($\mu^-$) in the dimuon rest frame with respect 
to the boost axis in the simplified models (I, II, and III), and 2HDM proposed in Ref.~\cite{Lane:2017gry}.
}
\end{figure}

\section{Kinematic distributions} 
\label{sec:kin}

In this section, we present a few kinematic distributions 
by making use of the parton level events that are generated by MadGraph5. Those distributions
could play  crucial roles 
in distinguishing  or excluding new physics models for the 30 GeV  dimuon excess. 
The kinematic distributions in Ref.~\cite{aleph30} include both signal and background, while the distributions in Fig.~\ref{fig:kin} contain only signal
from new physics models.
Thus, care should be exercised when we compare our predictions with the data.
However, we note that the continuum background events in Ref.~\cite{aleph30} are
just about half of all data in the signal region, our predictions for 
the kinematic distributions are not diluted. In addition, some kinematic
distributions in new physics models can definitely be distinguished
from the background as we will show later.

Among many of the kinematic variables studied in Ref.~\cite{aleph30}, the following four are chosen for representation and comparison purpose in this work (We denote the two muons and two b-jets of each signal events as $\mu_i, ~i=1,2$ and $b_i, ~i=1,2$, respectively.): 
\begin{itemize}
\item min angle($\mu ,b$) $\equiv$ angle($\mu_{i_0}, b_{j_0}$) $\equiv$ $\min_{i,j} \text{angle}(\mu_i, b_j)$,
\item the other muon angle($\mu ,b$) $\equiv$ $\min_{i,j} \text{angle}(\mu_i, b_j)$, where $i\neq i_0 $,
\item $p^{\text{rel}}_T(\text{closest pair}) \equiv |p(\mu_{i_0})| \times  \sin (\text{angle}(\mu_{i_0}, b_{j_0}) )$,
\item $\cos \theta^* (\mu^-)$, which is the angle of the muon in the dimuon rest frame with respect to the boost axis. 
\end{itemize} 

The distributions of those four variables are depicted in Fig.~\ref{fig:kin}. 
In the left top panel, we show the minimum angle($\mu_{i_0}, b_{j_0}$)
between a muon and a leading
jet.
For the benchmark points I (Model--I) and III (Model--II)
and for $m_\phi=50$ or $80$ GeV in the Model--III, we can see peaks between
$50\degree$ to $70\degree$ in the top left panel of Fig.~\ref{fig:kin}. 
However, for $m_\phi=10$ GeV in the Model--III,  the peak appears around $130\degree$.
The difference of two cases mainly comes from kinematics.
In the latter case, the muon pair is produced from the on-shell $\phi$
because $\phi$ is very light. Then the muon pair and two $b$-jets would
be produced back-to-back so that the direction of both muons would be opposite
to that of the leading jet. However, in the former case, the muon pair is 
produced in the off-shell 3-body decay so that the distribution could be milder
and the peak is shifted to the lower angle.

A similar thing happens in the other muon angle($\mu,b$) as shown
in the top right panel of Fig.~\ref{fig:kin}.
In the former case, the peak of this angle appears around $100\degree$,
while in the latter case, the peak around $150\degree$.
We note that the angle of the other pair of the muon and jet, which do not
take part in the minimum angle of a muon and a leading jet, has similar 
distribution to the top right panel of Fig.~\ref{fig:kin}, but
the peaks are slightly shifted to larger angles. 

In the bottom left panel of Fig.~\ref{fig:kin}, the relative transverse momentum
of the closest muon-jet pair has a peak at $16$ GeV in the latter case,
while in the former case it has much broader distribution and its maximum values 
can reach about $30$ GeV.

In the bottom right panel, we show the distributions of 
the decay angle $\cos\theta^*$ distribution for muons in the dimuon rest frame with 
respect to the boost axis  for each model. We find that only the benchmark point III (Model--II) shows 
the $\cos\theta^*$ distributions closer to the data, but not quite. 

In summary, we find that all the three models we considered in this paper have difficulty to  
accommodate the kinematic distributions.   Especially the third model discussed in Sec.~\ref{sec:model3} could
accommodate the rate without conflicting with other present data, but not the kinematic distributions.   
Note that we have exhausted all the models  
\footnote{In this paper, we do not consider 
$Z\rightarrow B \bar{B} \rightarrow (b \mu) (b\mu)$, 
since the existing lower  bound on $m_B$ indicates that one cannot reproduce 
$B(Z \rightarrow b\bar{b} \mu \mu ) \sim 10^{-5}$.} 
generating three topologically distinct Feynman diagrams for $Z\rightarrow b\bar{b} \mu\mu$ at tree level.
And we do not find any model could fit the kinematic distributions correctly.  
Since the ALEPH data seem to indicate that muons for the excess are likely produced with similar 
directions to the $b$ jets,  some muons might be from semileptonic $b$ decays.  
In order to understand this incompatibility of the kinematic distributions presented in Ref.~\cite{aleph30},
more detailed study of ALEPH data as well as other data on the $Z$ decays
may be necessary.

\section{Conclusions}
\label{sec:con}

In this letter we considered three different types of simplified models for the ALEPH 30 GeV
dimuon excess in $Z \rightarrow b\bar{b} \mu^+ \mu^-$.  The first class of models where 
a new resonance couples to $b\bar{b}$ and $\mu^+\mu^-$ are basically ruled out by the DY 
production of dimuon through $b\bar{b} \rightarrow X \rightarrow \mu^+ \mu^-$.  

In order to avoid the strong constraint from this DY process,  we considered the second model 
where  the 30 GeV dimuon excess is a spin-1 vector boson $Z'$ and proposed a new vectorlike singlet 
quark $B$, which has nonzero couplings for $Z$-$b$-$B$ and $Z'$-$b$-$B$.   Then we could account for the ALEPH
data without conflict with the DY constraint.  One can test this model at the LHC by $BB$,
$Bb$ and $Bj$ (with $j\neq b$) productions.   The subsequent decay of $B$ quark will result in 
the following final states: $bb + 4 \mu$, $b j + 2 \mu$ and $bb + 2\mu$, respectively, with
$O({\rm nb}), O({\rm pb})$, as summarized in Tables III and IV. These $B$-quark production cross sections
are sensitive to the $B$-quark mass $m_B$, and the current/future experimental studies of these 
channels will shed light on this class of models.  
Since the multi-muon final states have low background and
rarely seen at the LHC, this scenario is likely to be excluded already although there is no explicit search for
these final states.

Finally we considered a new $U(1)'$ gauge symmetry which is spontaneously broken by a nonzero VEV 
of a 
singlet scalar $\phi$ which has nonzero  $U(1)_Y$ and $U(1)'$ charges.
Then there appears a nonzero 
vertex for $Z$-$Z'$-$\phi$ which can accommodate Eq.~(\ref{brbbmm}) through $Z\rightarrow Z' \phi \rightarrow bb\mu\mu$.  
A natural choice for $U(1)'$ would be $U(1)_{L_\mu - L_\tau}$ gauge symmetry. 
For both simplified models II and III, the model building issue will be nontrivial, since  we need to introduce a 
new gauge symmetry associated with $Z'$, most likely with flavor dependence, and the simplest gauge 
symmetry might be  $U(1)_{L_\mu - L_\tau}$ gauge symmetry.   
In the model III,
we considered the case where $\phi$ 
is an $SU(2)_L$ singlet case, and found that the $U(1)'$ charge of $\phi$ should be very large $\sim O(10^4)$ (or $O(10^2)$ for the model with the kinetic mixing). 
The case $\phi$ being an $SU(2)_L$ doublet or higher representation is beyond the scope of this letter.   
We hope to address it in separate publications in the near future.

We also obtained some kinematic distributions in all the three models.
We find that the kinematic distributions in the three models
are not consistent with the ALEPH data. 
In this letter, we considered all possible scenarios to interpret the muon
excess as the decay of a resonance, but
the kinematic distributions in the ALEPH data might imply that
the muon excess is not likely due to the decay of a resonance.

\vspace{.5cm}
{\it Note Added:} 
While this paper was being reviewed, there appeared a paper which considers $Z\rightarrow H A$ 
in a certain type of 2HDM~\cite{Lane:2017gry}.  In Fig.~\ref{fig:kin}, we included the kinematic distributions 
in this model too, and conclude that the predictions from this new model are not consistent with the 
data~\cite{aleph30} either. 

\begin{acknowledgments}
We are grateful to Suyong Choi, Jack Kai-Feng Chen, Philip Coleman Harris and K.C. Kong for useful 
comments  on the subject presented in this letter.
This work is supported in part by National Research Foundation of Korea (NRF) Research 
Grant NRF-2015R1A2A1A05001869 (PK, JL, CY), and by the NRF grant funded by the Korea
government (MSIP) (No. 2009-0083526) through Korea Neutrino Research Center at Seoul 
National University (PK).
The work of CY is supported in part by the Do-Yak project of NRF under Contract
No. NRF-2015R1A2A1A15054533.

\end{acknowledgments}


\end{document}